# A CATALOG OF KAZARIAN GALAXIES

M.A. Kazarian[1], V.Zh. Adibekyan[1,2], B. McLean[3], R.J. Allen[3], A.R. Petrosian[2]

The entire Kazarian galaxies (KG) catalog is presented which combines extensive new measurements of their optical parameters with a literature and database search. The measurements were made using images extracted from the STScI Digitized Sky Survey (DSS) of $J_{pg}$(blue), $F_{pg}$(red) and $I_{pg}$(NIR) band photographic sky survey plates obtained by the Palomar and UK Schmidt telescopes. We provide accurate coordinates, morphological type, spectral and activity classes, blue apparent diameters, axial ratios, position angles, red, blue and NIR apparent magnitudes, as well as counts of neighboring objects in a circle of radius 50 kpc from centers of KG. Special attention was paid to the individual descriptions of the galaxies in the original Kazarian lists, which clarified many cases of misidentifications of the objects, particularly among interacting systems. The total number of individual Kazarian objects in the database is now 706. We also include the redshifts which are now available for 404 galaxies and the 2MASS infrared magnitudes for 598 KG. The database also includes extensive notes, which summarize information about the membership of KG in different systems of galaxies, and about revised activity classes and redshifts. An atlas of several interesting subclasses of KG is also presented.

## 1. Introduction

V.A.Ambartsumian's concept about the important role of activity in the nuclei of galaxies influencing their life and evolution [1] has been the stimulus for an extensive study of galaxies with active

[1] Yerevan State University, Yerevan, Armenia, e-mail: adibekyan@bao.sci.am
[2] V.A.Ambartsumian Byurakan Astrophysical Observatory, Byurakan 0213, Armenia
[3] Space Telescope Science Institute, 3700 San Martin Drive, Baltimore, MD 21218, USA

nuclei. The observations often showed that the spectra of galaxies with active nuclei have UV-excess radiation, since this phenomenon also is the characteristic of the activity.

For detecting new galaxies with UV-excess, two systematic 1.5° objective prism surveys were performed using the 1m Schmidt telescope of the Byurakan observatory by B.E.Markarian [2] and M.A.Kazarian [3].

The sky areas which were covered in the two surveys didn't overlap each other. The Markarian survey was completed in 1978 and published in a series of 15 papers including 1500 UV-continuum galaxies [4]. The catalogs of Markarian galaxies were published in 1986 [5], 1989 [6] and the latest optical database for the all Markarian galaxies published in 2007 [7]. The Kazarian survey started in 1970 and was finished in 1976. The survey results were published in a series of 6 papers including 702 UV-excess galaxies [3, 8-12] first of them was published in 1979.

Detailed spectrophotometric and morphological investigations of these galaxies with UV-excess have been done before publications of the papers [3, 8-12]. The first observations of these galaxies were made in 1973 with the Palomar 5m and the MacDonald 2.7m telescopes by E.Ye. Khachikian. The results of these observations were published in [13-17].

The systematic spectral and morphological observations of KG from survey [3, 8-12] began in 1978 by M.A.Kazarian with the 2.6m telescope of the Byurakan Astrophysical observatory (BAO), and the 6m telescope of the Special Astrophysical observatory (SAO) of the Russian Academy of Sciences. The last spectra of this series were observed on the 2.6m telescope with VAGR multi-pupil spectrograph for the galaxies Kaz 88, Kaz 138 and Kaz 146 in 2007. During the past 30 years, the spectroscopic observations for 156 KG have been made by the scientific group of the chair of astrophysics of Yerevan State University (YSU). The results of these observations have been published in about 60 articles. Among these galaxies have been discovered 17 Seyfert type galaxies: Kaz 17, Kaz 33, Kaz 73, Kaz 82, Kaz 102, Kaz 147, Kaz 151, Kaz 153, Kaz 163, Kaz 199, Kaz 214, Kaz 238, Kaz 243, Kaz 246, Kaz 323, Kaz 336 and Kaz 357, which composes about 10.9 % of the above mentioned 156 galaxies. The results of these Seyfert galaxies were published in the articles [17-38]. From these works it also became clear that many of these galaxies show very interesting physical characteristics. For example, Seyfert galaxy Kaz 17 is distinctive among the other KG by the quantity of emission and in particular FeI absorption lines [34,36]. The X-ray source Kaz 102 (QSO) shows dramatic and unexplained variability in different ranges of its spectrum [14,17,35,39-41], and Kaz 163 also shows strong variability in the spectrum [18,19,21,27,29,42] etc..

KGs have a wide range of physical parameters. Their linear diameters cover broad range of values from a few kpc (BCDs) up to a hundred kpc; they span a wide luminosity range between -23 and -13 mag and have a very broad range of morphological structure as well as different stages of activity from QSOs to BCDs. Therefore, the study of KG as a group, and the statistical comparison of their properties with a control sample of "normal field galaxies", and similar samples such as Markarian galaxies and KUGs can be used to understand a number of scientific problems. In [7], extensive lists of scientific problems that can be studied using the various parameters of UV-excess galaxies are listed.

So far no articles have been published with a complete summary of data for all KG. Today, the availability of the high-quality observations from photographic sky surveys allows us to measure additional optical parameters such as morphology, apparent magnitude, size and axial ratio of the galaxies in a more accurate and homogeneous way and also to extract quantitative data about their local environment. In this paper we describe the properties of the KG.

This paper describes a new database for 706 KG containing the following new measurements: accurate optical positions, morphological classes, apparent magnitudes (red, blue and near-IR), diameters, axial ratios and position angles, and counts of neighbour galaxies within 50 kpc radius based on the galaxy redshift and assuming a value for the Hubble constant of $H_0$ = 75 km s$^{-1}$ Mpc$^{-1}$. We also provide extensive notes indicating isolation or the membership of KG in groups, triplets, or pairs of galaxies. In addition, the objects with ambiguous activity classes or redshift determinations are included and described in the notes to the database. We have also included the corresponding near-IR magnitudes for all the galaxies from the 2MASS surveys. This will facilitate the comparison of the optical and near-IR properties of galaxy subsets.

In section 2 of this paper we describe the Kazarian survey and available informational data. The observational material used and the generation of the database are described in section 3. The database itself and related notes are described in section 4. In section 5 we present an atlas of several interesting subclasses of KG.

## 2. The Kazarian survey and available informational data

The Kazarian survey of UV excess galaxies was started in 1970. Observations were carried out with the Byurakan Observatory 1m Schmidt telescope equipped with a low-dispersion (2500 Å/mm at Hβ), 1.5° objective prism. This objective prism was used with Kodak 103a-E, Kodak IIa-E, Kodak II-AF and Kodak

IIa-F plates. The survey consisted of 88 fields (each 17 deg$^2$ in size) and covered about 1500 deg$^2$. The fields have been chosen so that galactic latitudes of their centers would be |b|>20°. The limiting magnitudes of these filters vary from 16 to 18 mags. The Kazarian survey was completed in 1976, and was published in a series of 6 papers including 702 UV-excess galaxies [3, 8-12].

Several KG have more than one component (Kaz 72, Kaz 267, Kaz 457 and Kaz 523) described as a UV-excess object. We distinguish all these components as individual KG and include them in this database by adding letters (a, b) to the name of the main Kazarian object. This brings the total number of Kazarian objects that are included in the present database to 706.

In the database, we include the spectral morphological characteristics (SMC) of galaxies, which have been taken from [3, 8-12]. The symbols s "stellar", d "diffuse" and also intermediate sd and ds were used to describe the morphology of spectra. For the degree of the UV emission the numbers 1, 2 and 3 were used. The detailed definition of these parameters is described in [43].

The redshifts for 404 KG have been collected from the literature. These include the NASA Extragalactic Database (NED), HYPERLEDA (Leon-Meudon Extragalactic Database), and the 64 KG observed spectroscopically in the SDSS 6$^{th}$ Data Release. In many cases when more than one redshift measurements are available, the more accurate HI 21 cm or the latest published value is given. References for redshift determinations for those galaxies for which the values of their redshifts differ more than 200 km/s are provided in the notes of the database.

We use the following classes to describe the activity: Seyfert class 1-2, Sy as unclassified Seyferts, LINERs as Seyfert class 3, QSO, BL Lacertae, SB (starburst nuclei), HII objects (spectra similar to HII regions), as well as AGN. If there is not sufficient spectral information for the classification of the galaxies, descriptions of the spectra of the galaxy as e (emission), a (absorption), or ea (emission, absorption) are presented. If the available activity classifications for the galaxy differ, the classification from the Veron-Cetty & Veron catalogue (2006) [44] or NED data is given (and these galaxies are mentioned in the notes to the database).

## 3. Measured parameters for Kazarian Galaxies

**3.1. Observational material and images of Kazarian Galaxies.** The optical measurements of these galaxies are based on the digitized blue, red and near-IR band images extracted from the photographic plates obtained by the second Palomar Observatory Sky Survey (POSS-II, [45]) and UKSTU surveys. All the plates were digitized at the Space Telescope Science Institute (STScI) using Perkin-Elmer PDS 2020G scanning microdensitometers with various modifications as described by [46] and were scanned at a resolution of 1.0" pix$^{-1}$. Approximately 93 % of the KGs are located in the northern hemisphere.

**3.2 The Coordinates.** The positional accuracy for KG presented in the first 5 original Kazarian lists are about 0.$^m$1 in right ascension and about 1' in declination [3], and for galaxies presented in the last Kazarian list the accuracy is about 2-3" [12].

In this paper, the galaxy coordinates were measured from the POSS-II *F* band plates. We visualized the images of the KG using the Aladin interactive software, and measured carefully the positions (using the peak intensity). In this case the positional uncertainty may be about 1.0". Note that multiple components have been renamed by adding a letter to the original Kazarian name. All coordinates are in the HST Guide Star Selection J2000.0 System.

**3.3 The Morphology.** The primary morphological descriptions of KG were presented in the original lists of KG [3, 8-12]. However, these descriptions were often in error because they were limited by the sensitivity of the Palomar Observatory Sky Survey (POSS-I) prints on which galaxies images were inspected. Note that in [47], the morphological descriptions for 141 KG based on the 2.6m BAO and 6m SAO telescopes observational data are presented. The current database presents a complete and homogeneous summary for the morphologies of KG.

For the morphological study of KG the $J_{pg}$(blue), $F_{pg}$(red) and sometimes $I_{pg}$(NIR) plates were used. By using both digital images and isophotal maps, which were constructed to display the large dynamic range of the images, we classified the KG using the modified Hubble sequence (E-S0-Sa-Sb-Sc-Sd-Sm-Im). 15 more galaxies in the sample have been classified as Compact and 25 galaxies as Close Interacting/Interacting types. In SDSS DR6 [48] there are images of 135 KG, for these galaxies we checked our classifications using SDSS g(4686 Å) and r(6165 Å) images.

Prior to this study, the HYPERLEDA, database contained the published morphological classes of 554 KG from which 318 are classified as S? and 236 are classified more fully. These were collected from various publications and are often incorrectly defined or not complete. Figure 1 illustrates a few examples that are misclassified in the HYPERLEDA compilation due to their inhomogeneous nature. For the above-mentioned 236 galaxies, in 22 cases the classification in HYPERLEDA are very different from ours (the difference > 5 units), for 52 galaxies the difference is not significant (the difference = 1,2 units) and for 143 galaxies the classifications are the same.

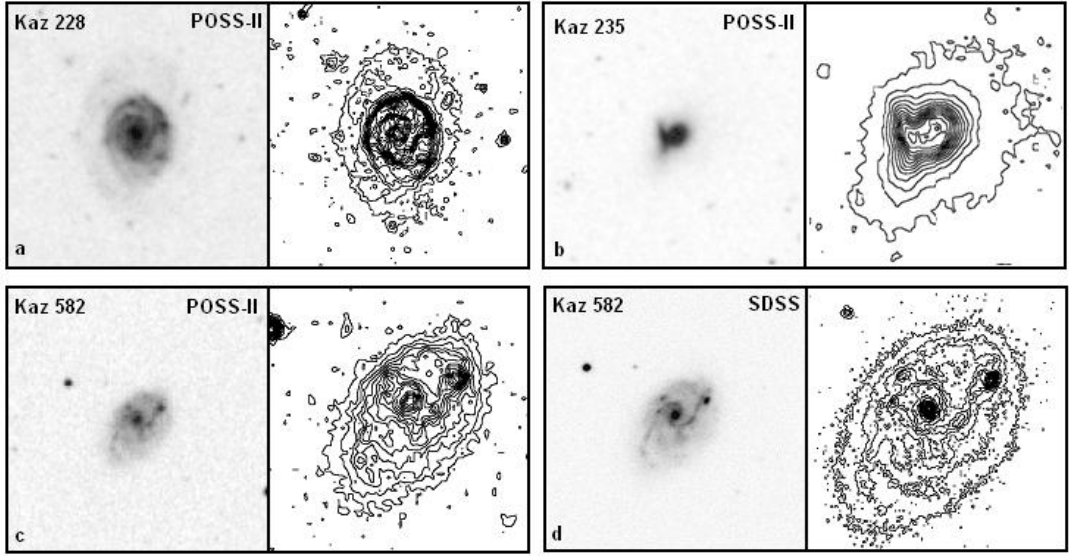

Fig. 1. - Examples of galaxies classified differently by HYPERLEDA and in this paper (North is up, and east is to the left). The contour levels are in arbitrary units. The lowest contour level was chosen at about 3σ level of the local background. The interval was chosen in order to best illustrate both inner and outer structure of the galaxy. Kaz 228 (a) is classified as E/S0 in HYPERLEDA and our classification is SBbc, Kaz 235 (b) – Sb by HYPERLEDA and – interacting in this paper and Kaz 582 (c) – ImB and SBc is our classification. In (d) the SDSS image of Kaz 582 is presented which also confirms our classification.

Figure 2 illustrates the distribution of the object morphologies derived from our homogeneous classification compared with the subsample available within HYPERLEDA.

**3.4 Apparent $J_{pg}$ magnitudes and $J_{pg}$-$F_{pg}$ and $F_{pg}$-$I_{pg}$ colors.** In this database we provide our measurements of the apparent blue, red and NIR magnitudes for all KG with improved accuracy and in a homogeneous manner. The magnitudes of the galaxies were measured from the POSS-II and UKSTU photographic survey plates that are available at STScI and used for the construction of the GSC-II catalog [49]. The technique used for determining the galaxy magnitudes was the same as described in [7]. The blue, red and NIR apparent magnitudes of the sample galaxies were measured from the $J_{pg}$, $F_{pg}$ and $I_{pg}$ band images in a homogeneous way at roughly the isophote corresponding to 3 times the background rms noise which is approximately 25.3 mag arcsec$^{-2}$ [7].

A comparison of our $J_{pg}$ band magnitude measurements with HYPERLEDA $B$-band magnitude determinations was also conducted for the KG. In this comparison we do not include only multi-component KG. Figure 3a compares HYPERLEDA B and our $J_{pg}$ magnitudes for 600 KG. The calculated transformations are as follows:

$$B = (1.0008 \pm 0.019)J_{pg} + (0.268 \pm 0.298), \text{ r} = 0.906 \pm 0.017, \text{ N} = 600 \qquad (1)$$

The mean absolute difference of the HYPERLEDA $B$ and our $J_{pg}$ magnitudes for KG is 0.37 ± 0.41. For 76 % KG the HYPERLEDA B and our $J_{pg}$ magnitude difference is less than 0.5 mag and, only for 8 % galaxies this difference is greater than 1.0 mag. The 0.2-0.5 mag internal error of our measurements has its contribution on individual differences, but the main reason is the 0.3 mag difference in the lowest isophotal level in HYPERLEDA and our system, which can have significant effect on the integrated blue magnitudes. For example Kaz 232, 429 and 569 for which magnitude differences are $-2^m.15$, $-4^m$ and $-2.^m66$ respectively. For these galaxies our blue diameters are also larger than HYPERLEDA blue (D25) diameters. The presence of close neighbors or projected stars also can introduce a larger error in either our or HYPERLEDA magnitude determinations, if they are not properly separated by the software algorithms.

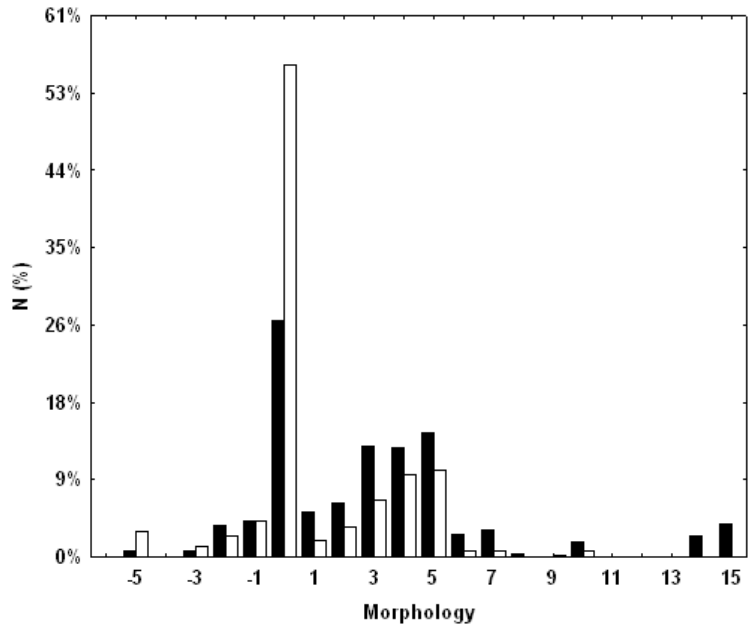

Fig. 2. – Bar graph (in percent) showing change in morphological distribution between the HYPERLEDA sample (white) and our complete classification (black).

**3.5 Angular Diameters, Axial Ratios and Position Angles.** In the original lists of KG the angular sizes of UV-excess galaxies were commonly measured on red POSS-I charts, and in some cases from the images of the galaxies observed by the Byurakan observatory 2.6m telescope and by the SAO 6m telescope. These diameters were eye estimates; hence they were not homogeneous or accurate.

The geometry (major and minor angular diameters, axial ratios and position angles) of KG was measured in a homogeneous way from the blue images of the galaxies at the same (25.3 mag arcsec$^{-2}$) isophotal level as the magnitudes measurements. The P.A. is measured from the north (P.A. = 0°) toward east between 0° and 180°.

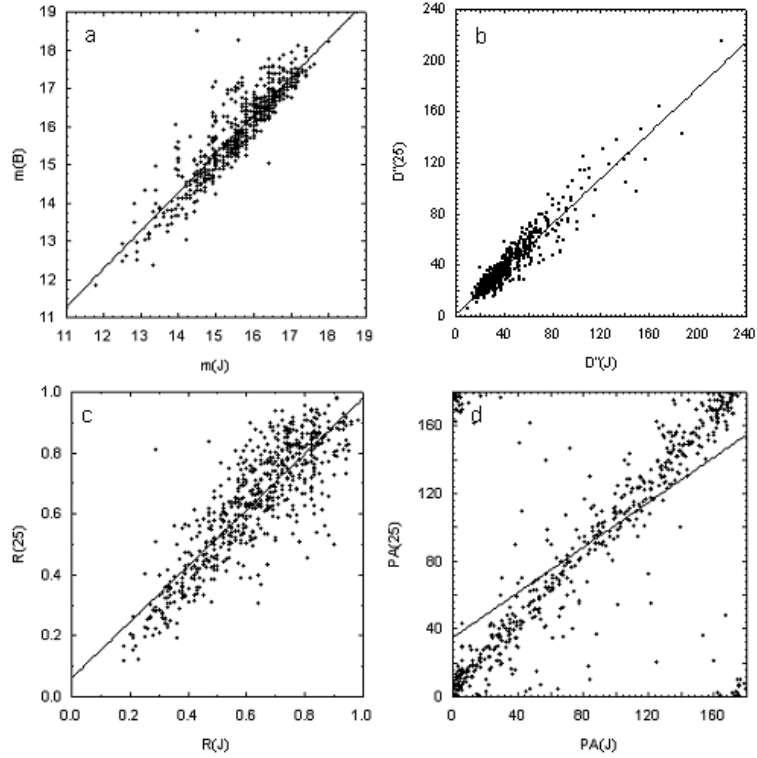

Fig. 3. – Comparison of the measured optical parameters with HYPERLEDA. The best linear fit lines are plotted.

A comparison of our blue angular diameters, axial ratios and position angles for KG with the HYPERLEDA determinations for the same objects has been done and it is shown in Figure 3 b,c and d respectively. The dependence between our and HYPERLEDA measurements can be fitted with the linear regression, which have the following forms and coefficients of correlation:

D"(25) = (0.884 ± 0.012)D"(J) + (1.956 ± 0.585), r = 0.947 ± 0.012, N = 629     (2)

R(25) = (0.919 ± 0.022)R(J) + (0.061 ± 0.014), r = 0.850 ± 0.021, N = 625     (3)

P.A.(25) = (0.666 ± 0.034)P.A.(J) - (34.900 ± 3.416), r = 0.640 ± 0.032, N = 548     (4)

Equation (2) and fig 3b show that our measured blue diameters are typically larger than HYPERLEDA blue D(25) diameters. This is due to the deeper mean limiting galaxy surface brightness in our system.

Fig 3c and (3) indicate that the KG axial ratios in HYPERLEDA and our measurements have no significant differences.

The mean absolute difference between the HYPERLEDA's and our blue P.A. for KG is $12.0^o \pm 12.6^o$. For 63 % KG the HYPERLEDA's and our blue P.A. difference is less than $10^o$ and for 87 % is less than $20^o$. But from fig. 3d it is visible that there is a large scatter, the possible reasons for which are the same as noted above (§3.4).

**3.6 Counts of Neighbor Galaxies.** Counts of neighboring galaxies were done for all galaxies which have determined redshifts z > 0.005 by projecting a circle of 50 kpc radius around the center each galaxy. All galaxies detected within this circle were counted if their angular sizes differed from that of the sample galaxy by no more than factor of 2 (e.g. [50]), and wherever redshifts were available, have a velocity difference within $\pm$ 800 km s$^{-1}$ (e.g. [51]). The counts of neighbor galaxies were checked in the 50 kpc circles extracted from both $J_{pg}$ and $F_{pg}$ band images. Objects closer than z=0.005 were not used because of the difficulty in reliably determining the associated objects over a wider field of view, as random projections become more dominant.

## 4. The Optical Database

A portion of the database for 10 KG is shown in Table 1 to illustrate its form and content. It contains observational data for 706 Kazarian objects with columns, described below. Table 1 is available in its entirety as the electronic variant of the catalog of KG in the VizieR Catalogue Service, via http://cdsarc.u-strasbg.fr/cgi-bin/VizieR?-source=VII/254.

Column (1). – Kazarian number as it appears in the original lists. For objects newly added to the database, we add an "a" or "b" letter designation (e.g. Kaz 72a and Kaz 72b). In all cases, the eastern galaxy or component is labeled "a" and the western galaxy or component with "b".

Column (2) and (3). - Equatorial coordinates (equinox J2000.0).

Column (4). - The morphological description of the galaxy. The following numerical codes were used: E = -5; E/S0 = -3; S0 = -2; S0/a = -1; S = 0; Sa = 1; Sab = 2; Sb = 3; Sbc = 4; Sc = 5; Scd = 6; Sd = 7; Sdm = 8; Sm = 9; Im = 10; Compact = 14 and Interacting system = 15. A bar is marked by "B".

Column (5). – SMC characteristics according to [3, 8-12]. For newly added KG SMCs are from descriptions of these objects in the original lists.

Column (6). – Activity class, when available. The various Seyfert classes are denoted by "Sy", "Sy1", "Sy1.5" and "Sy2". The symbol "Sy3" refers to LINERs. Starburst and HII nuclei respectively are indicated by "SB" and "HII". Galaxies with Wolf-Rayet futures in their spectra are indicated with the symbol "WR". In the table to describe the nuclei activity we use also symbols "AGN" and "NLAGN". QSOs and BL Lacertae objects are also identified. Also Kaz 348 galaxy is classified as LIRG (luminous infrared galaxy). Ambiguous activity classes are included and described in the corresponding notes to the database.

Column (7). - Heliocentric redshifts when available (literature, SDSS and NED). Again ambiguous redshifts data are included and described in the corresponding notes to the database.

Column (8-10). - Major diameter D(*J*) in arcseconds, Axial ratio R(*J*) and Position angle PA(*J*).

Column (11-13). – Apparent isophotal $J_{pg}$ magnitude and $J_{pg}$-$F_{pg}$ and $F_{pg}$-$I_{pg}$ colors.

Column (14). - Number of galaxies (Nn) detected within 50 kpc projected radius. The mark "nd" (no data) in this column relates to KG with redshifts smaller than 0.005 for which neighbor counts were not performed.

Column (15-17). - Near Infrared J magnitude and J–H and H–K colors, when available. These data are from 2MASS determinations [52]. Total J, H and K magnitudes of 2MASS are used to calculate these colors.

Column (18).—Attached notes: "s" denotes galaxies with information about their isolation or membership in the pair, triplet or group of galaxies; "a" denotes galaxies with ambiguous activity classification; "z" denotes galaxies with ambiguous determinations of their radial velocities. And also the symbol "*" has been used to describe any specifications of galaxies or if there was a comments. All notes and references to notes are available in the machine-readable table.

## 5. The Atlas

We also constructed a mosaic atlas for all KG where 2' x 2' regions of these galaxies from the digitized POSS-II $F_{pg}$ images are shown. Each plate shows 100 Kazarian objects. In Figure 4 we show the first 100 KG (the complete atlas is available, via ftp://cdsarc.u-strasbg.fr/pub/cats/VII/254/). The order within the plate is given by Kazarian numbering. The contrast of the images has been adjusted to provide the best subjective compromise between displaying the outer regions of the galaxies and preserving the structure of their inner regions.

As the connection between galaxy activity, star formation, and galaxy interactions is of especial interest, we pay special attention to the interacting KG. The close interacting systems are distinguished as a separate class of the objects according to the definitions presented by Petrosian et al. [7,53].

TABLE 1. Properties of Kazarian Galaxies

| Kaz | R.A. | Decl | Morph | SMC | AC | Z | D"(J) | R(J) | P.A.(J) | m(J) | $J_{pg}$-$F_{pg}$ | $F_{pg}$-$I_{pg}$ | Nn | J | J-H | H-K | Notes |
|---|---|---|---|---|---|---|---|---|---|---|---|---|---|---|---|---|---|
| (1) | (2) | (3) | (4) | (5) | (6) | (7) | (8) | (9) | (10) | (11) | (12) | (13) | (14) | (15) | (16) | (17) | (18) |
| 1 | 0 31 13.2 | -10 28 51 | 4 B | sd2 | HII | 0.0117 | 64 | 0.76 | 1 | 14.2 | 0.7 | 0.8 | 0 | 11.65 | 0.69 | 0.21 | s |
| 2 | 0 48 35.4 | -12 43 1 | 15 | ds1 | SB,WR | 0.0214 | 44 | 0.70 | 19 | 14.3 | 1 | 1.6 | 0 | 12.53 | 0.77 | 0.23 | s,* |
| 3 | 0 51 30 | -12 50 39 | 5 | ds2 | e | 0.0203 | 43 | 0.88 | 129 | 15 | 0.4 | 0.6 | 0 | - | - | - | - |
| 4 | 0 51 51.3 | -12 46 2 | 8 B | s1 | e | 0.0419 | 25 | 0.64 | 141 | 14.9 | 0.5 | 0.6 | 0 | 13.87 | 0.68 | 0.28 | - |
| 5 | 17 7 36.9 | +60 43 43 | 15 | ds1 | e | 0.0099 | 77 | 0.42 | 152 | 14.5 | 1.2 | 0.3 | 1 | 11.75 | 0.73 | 0.17 | s |
| 6 | 18 5 28.3 | +65 54 32 | 1 | s2 | e | 0.0282 | 28 | 0.77 | 40 | 16.3 | 1.8 | 0.7 | 1 | 13.25 | 0.63 | 0.38 | z |
| 7 | 0 3 3.9 | +34 22 41 | 0 | d3 | - | - | 22 | 0.75 | 47 | 16.4 | 1.5 | 0.8 | - | 14.44 | 0.24 | 0.51 | - |
| 8 | 0 3 35 | +23 12 3 | 5 | d2 | - | 0.0242 | 101 | 0.57 | 27 | 14.5 | 0.8 | 0.8 | 0 | 11.59 | 0.71 | 0.18 | - |
| 9 | 0 4 7.4 | +33 18 0 | 0 | s2 | - | - | 24 | 0.74 | 60 | 15.8 | 0.6 | 0.3 | - | 14.20 | 0.58 | 0.28 | - |
| 10 | 0 4 15.9 | +32 2 57 | 5 | d2 | - | - | 37 | 0.60 | 2 | 16 | 1.2 | 0.7 | - | 13.36 | 0.82 | 0.20 | - |

In Figure 5 we have shown grayscale representations and contour diagrams of the $F_{pg}$ band images of three galaxies in interaction systems. There are a total of 52 such systems of which 11 have both interacting objects as KG. The contour levels are in arbitrary units. The lowest contour level was chosen at the about 3σ level of the local background. The contour interval is constant, different in each case; usually it is between 10-30% of the local background. The scale interval was chosen in order to best illustrate both the inner and outer structure of a galaxy. In the same fashion field size (and thus magnification) was selected individually for each system to clearly illustrate its morphological structure. Note that this atlas does not include two interacting KG (Kaz 2 = Mrk 960 and Kaz 105 = Mrk 1100) that have been previously published [7]. The complete atlas of interacting systems of KG is available, via ftp://cdsarc.u-strasbg.fr/pub/cats/VII/254/).

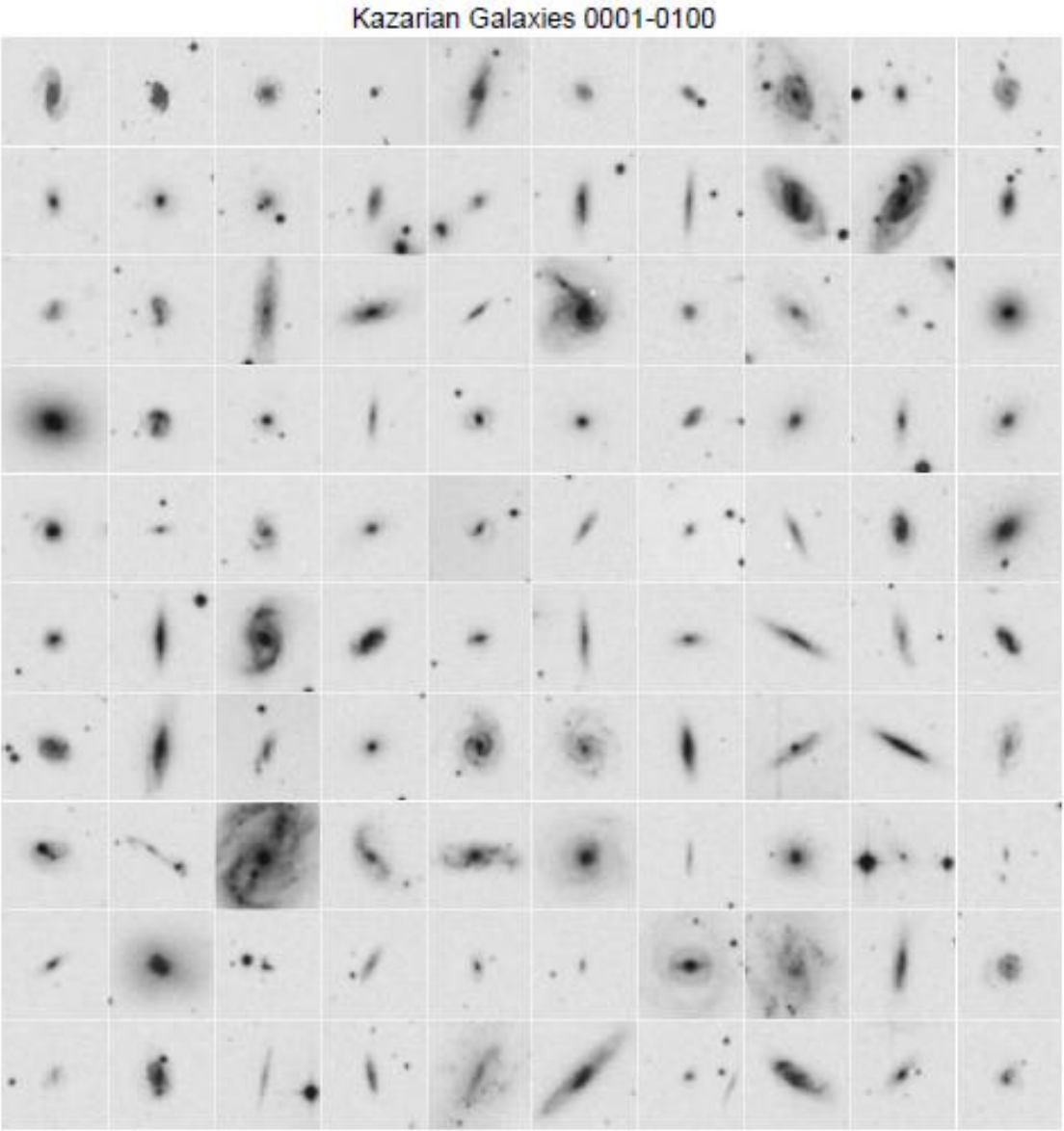

Fig. 4. — 2' x 2' field POSS-II $F_{pg}$ images for all KG. North is up, and east is to the left. The order within the plate is given by KG number.

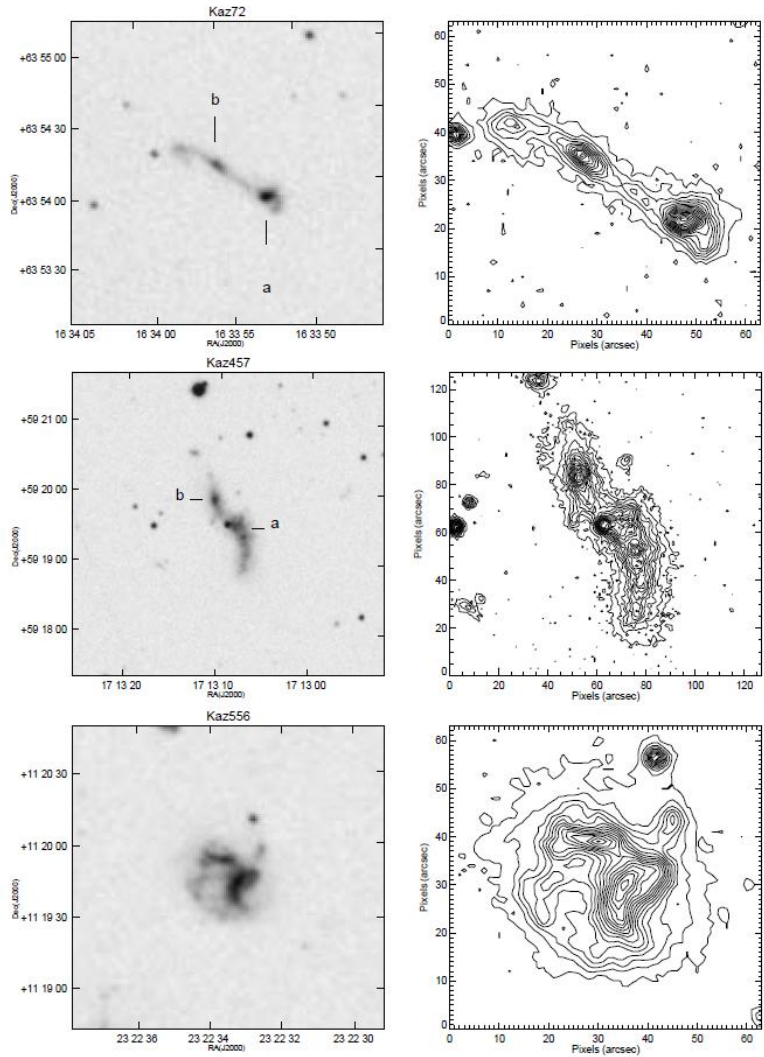

Fig. 5. — Gray-scale representations and contour diagrams of the $F_{pg}$-band images of KG in interacting systems. Contour levels and axis labels are as in Fig. 1.

**Acknowledgements**

This research has made use of NASA/IPAC Extragalactic Database (NED), which is operated by the Jet Propulsion Laboratory, California Institute of Technology, under contract with the National Aeronautics and Space Administration and HYPERLEDA (Leon-Meudon Extragalactic Database, http://cismbdm.univ-lyon1.fr/~hyperleda/). The Digitized Sky Survey was produced at the Space Telescope Science Institute under U.S. Government grant NAG W-2166. The images of this survey are based on photographic data obtained using the Oschin Schmidt Telescope on Palomar Mountain and the UK Schmidt Telescope. The plates were processed into the present digital form with the permission of these institutions. The Second Palomar Observatory Sky Survey (POSS-II) was made by the California Institute of Technology with funds from the National Science Foundation, the National Aeronautics and Space Administration, the National Geographic Society, the Sloan

Foundation, the Samuel Oschin Foundation, and the Eastman Kodak Corporation. The California Institute of technology and Palomar Observatory operate the Oschin Schmidt Telescope. Funding for the SDSS and SDSS-II was provided by the Alfred P. Sloan Foundation, the Participating Institutions, the National Science Foundation, the U.S. Department of Energy, the National Aeronautics and Space Administration, the Japanese Monbukagakusho, the Max Planck Society, and the Higher Education Funding Council for England. The SDSS was managed by the Astrophysical Research Consortium for the Participating Institutions. For image processing the ADHOC software (www.astrsp-mrs.fr/index_lam.html) developed by Dr. Jacques Boulesteix (boulesteix@observatoire.cnrs-mrs.fr; Marseille Observatory, France) was in intensive use.